\documentclass[aps,prl,twocolumn,showpacs,preprintnumbers,amsmath,amssymb,superscriptaddress,floatfix]{revtex4-1}

\usepackage{graphicx}% Include figure files
\usepackage{dcolumn}% Align table columns on decimal point
\usepackage{bm}% bold math
\usepackage{color}

\begin{document}

%\preprint{version 8}

\title{X-ray magnetic circular dichroism experiments and theory of transuranium Laves phase compounds}

\author{F. Wilhelm}
\affiliation{European Synchrotron Radiation Facility (ESRF), B.P.220, F-38043 Grenoble, France}
\author{R. Eloirdi}
\affiliation{European Commission, Joint Research Centre, Institute for Transuranium Elements, Postfach 2340, D-76125 Karlsruhe, Germany}
\author{J. Rusz}
\affiliation{Department of Physics and Astronomy, Uppsala University, Box 516, S-75120 Uppsala, Sweden}
\affiliation{Institute of Physics, Academy of Sciences of the Czech Republic, Na Slovance 2, CZ-182 21 Prague, Czech Republic}
\author{R. Springell}
\affiliation{European Synchrotron Radiation Facility (ESRF), B.P.220, F-38043 Grenoble, France}\affiliation{Royal Commission for the Exhibition of 1851 Research Fellow, Interface Analysis Centre, University of Bristol,  Bristol BS2 8BS, United Kingdom}
\author{E. Colineau}
\affiliation{European Commission, Joint Research Centre, Institute for Transuranium Elements, Postfach 2340, D-76125 Karlsruhe, Germany}
\author{J.-C. Griveau}
\affiliation{European Commission, Joint Research Centre, Institute for Transuranium Elements, Postfach 2340, D-76125 Karlsruhe, Germany}
\author{P. M. Oppeneer}
\affiliation{Department of Physics and Astronomy, Uppsala University, Box 516, S-75120 Uppsala, Sweden}
\author{R. Caciuffo}
\affiliation{European Commission, Joint Research Centre, Institute for Transuranium Elements, Postfach 2340, D-76125 Karlsruhe, Germany}
\author{A. Rogalev}
\affiliation{European Synchrotron Radiation Facility (ESRF), B.P.220, F-38043 Grenoble, France}
\author{G. H. Lander}
\affiliation{European Commission, Joint Research Centre, Institute for Transuranium Elements, Postfach 2340, D-76125 Karlsruhe, Germany}

\date{11/03/2013}
% It is always \today, today,%  but any date may be explicitly specified

\begin{abstract}
The actinide cubic Laves compounds NpAl$_{2}$, NpOs$_{2}$, NpFe$_{2}$, and PuFe$_{2}$ have been examined by X-ray magnetic circular dichroism (XMCD) at the actinide $M_{4,5}$ absorption edges and Os $L_{2,3}$ absorption edges. They have the interesting feature that the $An-An$ spacing is close to the so-called Hill limit so that substantial hybridization between the 5$f$ states on neighboring atoms is expected to occur. The XMCD experiments performed at the $M_{4,5}$ absorption edges of Np and Pu allow us to determine the spectroscopic branching ratio, which gives information on the coupling scheme in these materials. In all materials the intermediate coupling scheme is found appropriate. Comparison with the SQUID data for NpOs$_{2}$ and PuFe$_{2}$ allows a determination of the individual orbital and spin magnetic moments and the magnetic dipole contribution \textit{m}$_{md}$. The resulting orbital and spin magnetic moments are in good agreement with earlier values determined by neutron diffraction, and the values of \textit{m}$_{md}$ are non-negligible, and close to those predicted for intermediate coupling. There is a comparatively large induced moment on the Os atom in NpOs$_{2}$ such that the Os contribution to the total moment per formula unit is $\sim$30\% of the total. The spin and orbital moments at the Os site are parallel, in contrast to the anti-parallel configuration of Os impurities in 3$d$ ferromagnetic transition metals. Calculations using the LDA+\textit{U} technique are reported. The \textit{ab initio} computed XMCD spectra show good agreement with experimental spectra for small values (0-1eV) of the Hubbard \textit{U} parameter, which underpins that 5$f$ electrons in these compounds are relatively delocalized. The calculations confirm the sign and magnitude of the experimentally determined induced magnetic moments on the Os site in NpOs$_{2}$. \textit{A posteriori}, by comparison of the theoretical and measured XMCD spectra, we can determine the most appropriate LSDA+\textit{U} variant and the value of \textit{U}.
\end{abstract}

% PACS, the Physics and Astronomy Classification Scheme.
\maketitle

\section{I. Introduction}

The central question of the solid-state physics of the actinides is the extent and influence of the hybridization of the 5$f$ electrons. The vast panoply of properties, from localized to itinerant systems, from semiconductors to superconductors, is caused by variations of this property, as a consequence of the atomic configuration and environment. The 5$f$ electrons have an extended wavefunction in real space so that they can often interact with 5$f$ states of neighboring atoms; more often, however, hybridization occurs between the 5$f$ states and other electron states of the same or neighboring atoms. For example, in metallic systems the 5$f$'s almost always interact with the conduction states, which are made up of 6$d$ and 7$s$ states. In compounds with extended $p$ or $d$ states the 5$f$'s can have complicated interactions with such states, and a surprising number of different phenomena are observed in the solid-state properties\cite{moore09,santini,pfleiderer,mydosh}.

Many years ago Hill\cite{hill70} pointed out that there was a connection between the properties of actinide compounds depending on the inter-actinide spacing $d_{An}$ in the material. For small $d_{An}$ the materials tended to be superconductors, whereas for large $d_{An}$ the materials tend to order magnetically. These ideas addressed the hybridization between 5$f$ states on neighboring sites. In today's language we would recognize that for small $d_{An}$ the bandwidth is large, favoring a suppression of magnetism and possible superconductivity, whereas for large $d_{An}$ the bandwidth is small and there is a tendency to localization of the 5$f$ states, leading to magnetic order. Despite the Hill criterion now being 40 years old, the ideas are still of interest. The critical spacing where these phenomena change was estimated by Hill as $d_{An}$$\sim$3.3{\AA}.

It was recognized about this time that the actinide cubic Laves phases (C15 face-centered cubic structure of MgCu$_{2}$ type) were well-packed structures with $d_{An}$ close to the critical spacing. A systematic study of these materials\cite{lam72,harvey74,aldred74,aldred75a,aldred75b,aldred76,lander77,aldred79} in polycrystalline form was performed at Argonne National Laboratory in the 1970s. The measurements reported were electrical resistivity, magnetization, M\"{o}ssbauer spectroscopy, and neutron scattering. We shall draw on this body of work, but note that these techniques were unable to give any detailed information on the individual spin and orbital moments on the actinide (or other) ions. Since it is important to know these values to compare with theory, which is now capable of treating the strong hybridization observed, we have revisited some of these materials with x-ray magnetic circular dichroism (XMCD). XMCD is element and shell specific and can obtain the individual spin and orbital moments in the structure on different atoms.

In this paper we shall concentrate on the ferromagnetic systems, NpAl$_{2}$, NpOs$_{2}$, NpFe$_{2}$, and PuFe$_{2}$. Some of their properties already known are shown in Table \ref{tablemagpar}. We use for PuFe$_{2}$ the results from a 1988 study of a single crystal with polarized neutrons\cite{wulff88}.

\begin{table*}
\caption{ Compounds discussed in this paper. $a_{o}$ is the lattice parameter, $d_{An}$ is the actinide-actinide interatomic distance ($\sqrt{3}$/4)$a_{o}$, $T_{C}$ is the Curie temperature,  $\mu_{o}$ is the spontaneous ordered magnetic moment determined by magnetization studies,  $\mu_{An}^{N}$ is the magnetic moment at the actinide site determined by neutron diffraction, $\mu_{Np}^{Hf}$ is the magnetic moment determined at the Np site by assuming the hyperfine field is proportional to the moment\cite{dunlap74}, and $\mu_{T}$ is the magnetic moment at the $T$ site, determined in these cases by neutron diffraction. Recall that the Hill criteria\cite{hill70} gives the critical value of $d_{An}$$\sim$3.3{\AA}.}
\label{tablemagpar}
\begin{ruledtabular}
\begin{tabular}{cccccccc}
Compound& $a_{o}$({\AA})& $d_{An}({\AA})$& $T_{C}$(K)& $\mu_{o}$($\mu_{B}$/f.u.)&  $\mu_{An}^{N}$($\mu_{B}$)& $\mu_{Np}^{Hf}$($\mu_{B}$)& $\mu_{T}$($\mu_{B}$)\\
\hline\\
  NpAl$_{2}$ &	7.785 & 3.37 & 56(1)  & 1.21(1)  & 1.50(5) & 1.51(4) &  -- \\
  NpOs$_{2}$ &  7.528 &	3.26 & 7.5(5) & 0.44(3)  & 0.25(5) & 0.40(4) &   --\\
  NpFe$_{2}$ &  7.144 &	3.09 & 492(8) & 3.38(10) & 1.09(8) & 0.87(6) & 1.35(5)\\
  PuFe$_{2}$ &  7.150 &	3.10 & 564(8) & 3.4(2)   & 0.39(2) &   --    &  1.73(1)\\
\end{tabular}
\end{ruledtabular}
\end{table*}

Important conclusions from this body of work were:
\begin{enumerate}
\item  NpAl$_{2}$ is probably a localized system ($d_{An}$ is larger, the moment is larger, and $T_{C}$ higher, than in NpOs$_{2}$).
\item  NpOs$_{2}$ is probably an itinerant system, with a low moment and $T_{C}$ and almost negligible magnetic anisotropy.
\item  In the case of both NpFe$_{2}$ and PuFe$_{2}$ the actinide 5$f$ states are mostly itinerant.
\item  There is hybridization between the actinide 5$f$ and Fe 3$d$ states as shown by the polarized-neutron study\cite{wulff88} of PuFe$_{2}$ and partial quenching of the orbital moment at the Pu site.
\end{enumerate}

Starting in 1989 Eriksson, Johansson, and Brooks published a sequence of theory papers\cite{eriksson89,eriksson90a,eriksson90b,eriksson90c} addressing particularly the magnetic properties of these Laves phase actinide materials. These band-structure calculations were performed with the local-density approximation (LDA) and the large spin-orbit interaction was included. The results gave the ordered moments, as well as their constituent orbital ($\mu_{L}$) and spin ($\mu_{S}$) parts.  Since the magnetic form factor, as measured in neutron scattering, depends on ratio of the orbital to spin contributions\cite{lander91}, they also predicted the magnetic form factors. In 1993 the question of orbital polarization within the local-spin-density approximation (LSDA) was considered\cite{severin93} giving, by this time, a fairly complete theory.

Since that time, a new powerful magnetic characterization tool has emerged; the X-ray magnetic circular dichroism (XMCD) technique. This synchrotron based technique  allows one  to quantitatively estimate the spin and orbital magnetic moments of the absorbing atoms. XMCD has been demonstrated to work successfully for the 3$d$, 4$d$ and 5$d$ transition metals. Indeed, a set of sum-rules relate the experimental integrated XANES and XMCD spectra to the expectation value of ground state operators. More recently, this technique has been used extensively for a number of uranium compounds at the $M_{4,5}$ absorption edges ($M_{4}$ = 3.73keV; $M_{5}$ = 3.55keV) where the dipole-allowed signal corresponds to the promotion of an electron from a core 3$d$ shell to the partly filled 5$f$ band. Such experiments have been performed on ferromagnetic uranium compounds with the NaCl structure\cite{collins95,dalmasdereotier97,dalmasdereotier98,kernavanois01}, the Laves phase ferromagnet UFe$_{2}$\cite{finazzi97}, heavy-fermion materials where the moment is induced with a magnetic field\cite{yaouanc98,dalmasdereotier99}, UTAl ferromagnets\cite{kucera02}, and even multilayers containing uranium\cite{wilhelm07,springell08}. Up to now, only two XMCD studies  have been reported on the $M_{4,5}$ absorption edges of a transuranium material, NpNiGa$_{5}$, by Okane \textit{et al.} \cite{okane09} and work at the ESRF on Np$_{2}$Co$_{17}$\cite{halevy12}.

There have also been a number of theory efforts calculating the band structures and comparing the resulting XAS and XMCD signals with experiment\cite{shishidou99,antonov03a,antonov03b,yaresko03,antonov04,kunes01a}. Using atomic theory, Thole and Carra\cite{thole92,carra92,carra93} have derived sum-rules that relate integrated intensities of XAS and XMCD to the ground-state expectation values of standard operators, such as the orbital $\langle L_{z}\rangle$ and spin $\langle S_{z}\rangle$ magnetic moments of the absorbing atom. Moreover, atomic multiplet theory has been used to calculate the atomic core-level spectra at the $M_{4,5}$ absorption edges, which is associated with the transitions $f^{n} \rightarrow d^{9}f^{n+1}$ (\cite{vanderlaan96}, \cite{okane09}). However, there was no attempt to calculate the spectral shape of both the XAS and XMCD for neptunium using band-structure calculations.

XMCD can also be performed on the actinides at the $N_{4,5}$ edges, where the transitions are from the 4$d$ core shell to the partially filled 5$f$ shell and the transition energies are less than 1 keV; however, these experiments are performed in the soft x-ray range. Due to the large core-hole lifetime broadening at the N edges, it is more difficult to record an XMCD signal than with the $M$ edges. Furthermore, at these low excitation energy the experiments are extremely surface sensitive, so it is not certain that bulk-like properties will be measured. For a recent example see Okane \textit{et al.} in Ref.\cite{okane08}.

Although the XMCD technique is powerful, there are a few caveats to recall. The first is that although the orbital moment can be estimated  directly, the spin moment cannot. Rather than $\langle S_{z}\rangle$, the quantity deduced is $\langle S^{eff}\rangle$ where this is the effective spin moment. From the sum rules \cite{thole92,carra92,carra93}, we find that for the $M_{4,5}$ absorption edges,

\begin{equation}\label{seff}
\langle S^{eff}\rangle = \langle S_{z}\rangle + 3\langle T_{z}\rangle
\end{equation}

where $\langle T_{z}\rangle$ is the magnetic dipole operator, which describes correlations between the spin and position of each electron. $\langle T_{z}\rangle$ cannot be easily estimated experimentally, but it may be calculated assuming the electron count and the coupling of the electron states [$jj$, LS, or intermediate coupling (IC)]. Tabulated values are given in Ref.\cite{vanderlaan96}.  On the other hand, for actinide systems it is often these exact quantities (electron count and coupling scheme) that are unknown. We can determine $\langle T_{z}\rangle$, equivalently the magnetic dipole contribution \textit{m}$_{md}$=-6$\langle T_{z}\rangle$, by knowing precisely the magnetization since the total moment  $\mu_{tot}$ =  $\mu_{L}$ +  $\mu_{S}$, and combining this with the information from XMCD experiments. There is considerable interest in this quantity for the actinides as for localized uranium systems $\langle T_{z}\rangle$ appears to be large and close to the IC value \cite{dalmasdereotier97}, whereas the value of $\langle T_{z}\rangle$ seems to be close to zero for itinerant systems possessing a small magnetic anisotropy, such as UFe$_{2}$\cite{finazzi97}.

Determining the value of $\langle T_{z}\rangle$ by comparing the signal obtained in XMCD and the magnetization obtained by SQUID magnetometry may appear simple, but with the photon energies at the actinide $M$ (or especially with $N$) absorption edges the signal is sensitive to surface effects on the sample. Thus the SQUID is measuring a bulk signal, whereas the XMCD measures the signal from the first 200nm, or even less in the case of the $N$ absorption edges, of the sample. As we discuss below, it appears that magnetic domains near to the surface are strongly constrained.

The objective of the present work is to increase our knowledge of the AnX$_{2}$ systems shown in Table I. In particular, we wish to establish the orbital and spin moments experimentally, and determine the $\langle T_{z}\rangle$ for transuranium systems. Comparison with the values found from the latest electronic-structure theories tests our understanding of the physics of these compounds, the role of the 5$f$ electrons, as well as reveals which computational approaches provide the best description.

The plan of the paper is in Sec. II to give the experimental details, in Sec. III we present the experimental results, in Sec. IV we report the theoretical calculations and their results, and a discussion is presented in Sec. V.

\section{II. Experimental details}

The actinide Laves phases used in this study are all stoichiometric compounds and relatively easy to prepare in polycrystalline form by standard techniques.

The polycrystalline samples of NpAl$_{2}$, NpFe$_{2}$ and NpOs$_{2}$ were prepared by arc melting stoichiometric amounts of high-purity elemental constituents Np (99.9\% pure), Al (purity 5N), Fe(purity 5N) and Os (99.95\% pure) on a water-cooled copper hearth under an Ar (purity 6N) atmosphere. A Zr alloy was used as an oxygen getter. The samples were melted five times to improve the homogeneity and the mass loss was below 0.5\%.
Phase analysis of the obtained ingots was performed. Crystallographic analyses were performed at room temperature by x-ray diffraction (XRD) on samples with a mass of about 25mg, finely ground and dispersed on a Si wafer. Data were collected in back-reflection mode with a Bruker D8 diffractometer and Cu K$\alpha$ radiation selected by a Ge(111) monochromator. A one-dimensional position sensitive detector was used to cover the angular range from 15 to 120$^\circ$ with incremental steps of 0.0085$^\circ$. The analysis of the XRD pattern confirms a cubic C-15 Laves phase structure. In all cases x-ray diffraction was used to identify that single-phase material was present, with lattice parameters consistent with those reported in Table I. In the case of PuFe$_{2}$, the sample had been prepared previously at ITU and pieces of that preparation were used for the SQUID and XMCD experiments.

The NpAl$_{2}$, NpFe$_{2}$ and NpOs$_{2}$ samples were then polished to get platelets with two flat surfaces necessary for XMCD experiments and sealed in a specific capsule. The samples, which were between 10-20mg (except PuFe$_{2}$ which was around 0.2mg), were glued with epoxy resin in the middle of the Al support. A 100$\mu$m thick Be window, covering the sample, was then glued with stycast on the support. For safety reasons, the Be window was then covered with a 13$\mu$m thin kapton foil. Finally, the capsule was closed with an Al upper part, screwed on the support. All the encapsulation procedure was done under helium atmosphere, to avoid any surface oxidation. Those capsules are sealed at ambient pressure and sustain liquid helium temperature. From the NpOs$_{2}$ batch, we kept one sample piece as-grown, which we will label the as-cast NpOs$_{2}$ sample.

Magnetization and magnetic-susceptibility measurements were carried out in the temperature range between 2 and 300 K and in magnetic fields up to 7 T using a Quantum Design MPMS-7 superconducting-quantum-interference-device (SQUID) magnetometer.

The x-ray-absorption-spectroscopy (XAS) and XMCD experiments were carried out at the ID12 beamline of the European Synchrotron Radiation Facility (ESRF), which is dedicated to polarization-dependent spectroscopy in the photon-energy range from 2 to 15 keV\cite{rogalev01}. For the experiments at the $M_{4,5}$ absorption edges of Np and Pu (3.6-4.1 keV), the source was the helical undulator Helios-II, which provides a high flux of circularly-polarized x-ray photons with a polarization rate in excess of 0.95. After monochromatization with a double-crystal-Si(111), the rate of circular polarization is reduced to about 0.42 at the $M_{5}$ edge and 0.50 at the $M_{4}$ absorption edges for Np and to about 0.46 at the $M_{5}$ edge and 0.54 at the $M_{4}$ absorption edges for Pu.  The x-ray-absorption spectra were recorded using the total-fluorescence-yield detection mode in backscattering geometry for parallel $\mu^{+}(E)$ and antiparallel $\mu^{-}(E)$ alignments of the photon helicity with respect to a 17T external magnetic field applied along the beam direction. Element selective magnetization curves were recorded by monitoring the intensity of the XMCD signal at a given photon energy, as a function of the applied magnetic field.

The x-ray-absorption spectra for right and left circularly polarized x-ray beams were done assuming semi-infinite samples, but taking into account the various background contributions (fluorescence of subshells and matrix as well as coherent and incoherent scattering), the angle of incidence of the x-ray beam, and finally the solid angle of the detector\cite{goulon82,troger92,pfalzer99}.

The Np and Pu edge-jump intensity ratio $M_{5}/M_{4}$ (defined as the ratio between the occupation numbers for the two spin-orbit-split core levels $j=3/2$ and $5/2$) were then both normalized to 1.57 according to the theoretical edge-jump ratio tabulated in the XCOM tables by Berger \textit{et al.} in Ref. \cite{bergerxcom}\footnote{The edge jump ratio of the $M_5$ to $M_4$ edges can be deduced from the entries in these tables for all actinides. They are all 1.57$\pm0.05$, except for Pu, which is listed as 1.87. We believe this value is incorrect and we have used 1.57 for our data analysis}.

The XMCD spectra $\mu^{+}(E)- \mu^{-}(E)$ were obtained as the difference of the corrected x-ray-absorption spectra. To make sure that the final XMCD spectra are free of experimental artifacts, measurements were also performed for the opposite direction of the applied magnetic field.

\section{III. Experimental results}

\subsection{A. Macroscopic magnetic characterization}

\begin{figure}
\includegraphics[width=8.5cm]{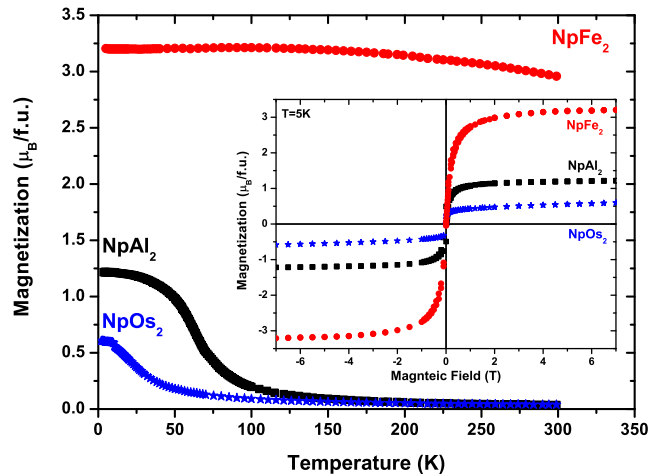}
\caption{(Color online) Temperature dependence of the magnetization measured on NpAl$_{2}$, NpFe$_{2}$ and NpOs$_{2}$ polycrystalline samples from 4 to 300 K in a field $\mu_{0}H$ = 7 T. The inset shows the magnetic field dependence of the magnetization measured on NpAl$_{2}$, NpFe$_{2}$ and NpOs$_{2}$ at 5 K.
\label{MagnetizationNp}}
\end{figure}

Fig. \ref{MagnetizationNp} shows the temperature ($T$) and magnetic field ($\mu_{0}H$) dependence of the magnetization obtained with the SQUID magnetometer for NpAl$_{2}$, NpFe$_{2}$, NpOs$_{2}$. The ordering temperatures and magnetization are similar to those reported by Aldred \textit{et al.}\cite{aldred74,aldred75a,aldred75b,aldred76}. Note in this respect that the value for Np compounds in Table \ref{tablemagpar} have been extrapolated to B = 0 T, whereas no attempt has been made to perform that extrapolation in the present work as we shall compare magnetization and SQUID data at B = 7 T. It was already noted that NpOs$_{2}$ has a large high-field susceptibility \cite{aldred76}. For the PuFe$_{2}$ sample, the ordering temperature and magnetization were the same as reported earlier \cite{lander77}.

\subsection{B. Absorption spectra and branching ratios at the actinide M edges}

Isotropic absorption spectra from the three NpX$_{2}$ samples are shown in Fig. \ref{XanesBR} (top panel) with that for PuFe$_{2}$ in the lower panel. Since the samples all have cubic symmetry, we can compare the x-ray absorption signals recorded at the $M_{4,5}$-edges directly. Clearly, the isotropic x-ray absorption spectra  for all three NpX$_{2}$ samples are similar. From the branching ratio, $B = I_{M_{5}}/(I_{M_{5}}+I_{M_{4}})$, where $I_{M_{4,5}}$ is the integrated intensity of the isotropic white lines at the $M_{4,5}$ edge, we may determine the expectation value of the angular part of the valence states spin-orbit operator, $\langle\psi|\vec{\ell}\cdot\vec{s}|\psi\rangle$ = $3/2\langle W^{110}\rangle$, as \cite{vanderlaan04}:

\begin{figure}
\includegraphics[width=8.5cm]{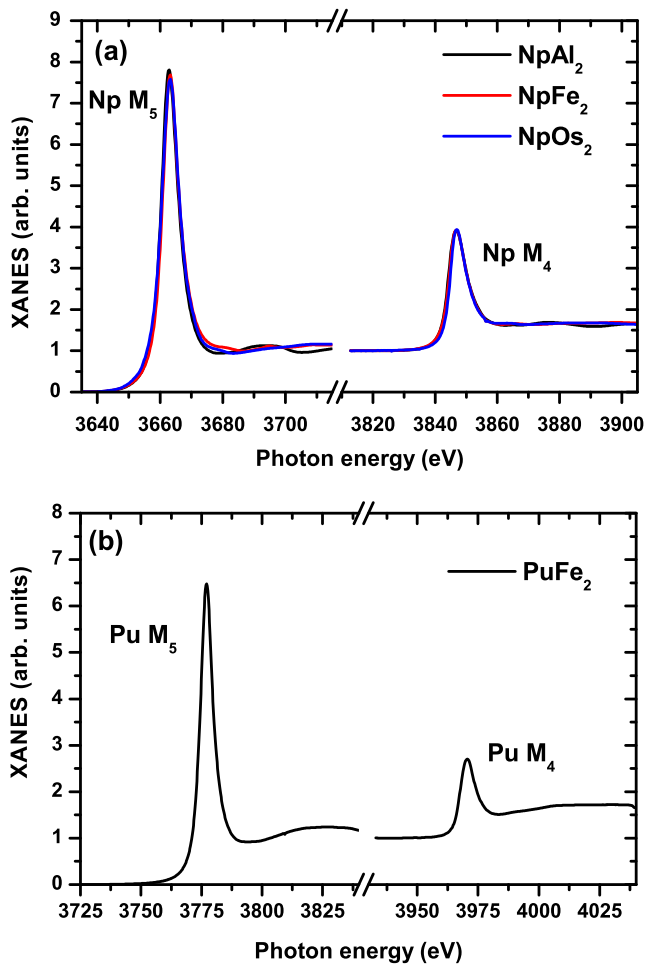}
\caption{(Color online) Isotropic x-ray absorption spectra recorded at the $M_{5}$ and $M_{4}$ Np edges for the polished NpX$_{2}$ samples (top panel (a)) and recorded at the Pu edges for the as-cast PuFe$_{2}$ sample (lower panel (b)) as a function of incident photon energy. The spectra have been corrected for self absorption effects.
\label{XanesBR}}
\end{figure}

\begin{equation}\label{so}
\frac{2\langle\vec{\ell}\cdot\vec{s}\rangle}{3n_{h}} = -\frac{5}{2}(B-\frac{3}{5})+ \Delta
\end{equation}

where $n_{h}$ is the number of holes in the 5$f$ shell, and $\Delta$ is a quantity dependent on the electronic configuration. Free-ion values for $\Delta$ have been calculated \cite{vanderlaan04}, and are -0.005 for 5$f^{4}$ and zero for 5$f^{5}$.

M\"{o}ssbauer spectroscopy on the NpX$_{2}$ materials\cite{aldred76} has shown that the ionic state of Np is close to 5$f^{4}$, and the neutron experiments reported on PuFe$_{2}$ \cite{wulff88} have shown that the ionic state of Pu is close to 5$f^{5}$. The corresponding number of 5$f$ holes are therefore 10 for the NpX$_{2}$ and 9 for PuFe$_{2}$. Density functional theory calculations (see Sec. IV) give values of 10.3 and 9.1 for actinides in NpOs$_{2}$ and PuFe$_{2}$, respectively, close to the expected ionic states. Moreover, since $\langle W^{110}\rangle$ = $n_{7/2}$ - 4/3 $n_{5/2}$, where the number of electrons in the individual shells corresponding to $j=l \pm s$, \textit{i.e.} $j=7/2$ and $j=5/2$ are $n_{7/2}$ and $n_{5/2}$, respectively, we may determine the occupation of the individual spin-orbit split electron shells \cite{vanderlaan04,moore09}. These quantities are presented in Table \ref{tableBR}.

\begin{table}
\caption{ Experimental branching ratio $B$ deduced from the x-ray absorption spectra, together with derived 5$f$ -electron contribution to the valence spin-orbit interaction per hole $\langle W^{110}\rangle$/$n_{h}$-$\Delta$;the electronic 5$f$-state occupations $n_{e}$ and the electron occupation of the $j=5/2$ and $j=7/2$ sub-shells. Experimental error bars are $\sim\pm1\%$ }
\label{tableBR}
\begin{ruledtabular}
\begin{tabular}{cccccc}
Compound& $B$& $\langle W^{110}\rangle$/$n_{h}$-$\Delta$ & $n_{e}^{5f}$& $n_{5/2}^{5f}$&  $n_{7/2}^{5f}$ \\
\hline\\
  NpAl$_{2}$ &	0.738 & -0.345 &  4  & 3.21  & 0.79 \\
  NpFe$_{2}$ &  0.742 &	-0.355 &  4  & 3.26  & 0.74 \\
  NpOs$_{2}$ &  0.750 &	-0.375 &  4  & 3.34  & 0.66 \\
  PuFe$_{2}$ &  0.803 &	-0.507 &  5  & 4.10  & 0.90 \\
\end{tabular}
\end{ruledtabular}
\end{table}

From Table  \ref{tableBR}, we can observe that the branching ratio $B$, provided by the XAS experiment, for Np in the various NpX$_{2}$ compounds are all similar and within the error bar is equal to that for Np metal ones which is 0.74 \cite{moore09}. Further, no difference has been observed between the as-cast and polished NpOs$_{2}$ samples. One obtains a 5$f$ spin-orbit interaction per hole of  about -0.36, very close to the value calculated in the intermediate-coupling approximation for the 5$f^{4}$ electronic configuration \cite{vanderlaan04}. This supports the conclusion drawn from the M\"{o}ssbauer isomer shift. In the case of PuFe$_{2}$, the branching ratio is about 0.80 which is close to the Pu metal value of 0.82 \cite{moore09}. Furthermore, it agrees well with the value calculated in the intermediate-coupling approximation for the 5$f^{5}$ electronic configuration \cite{vanderlaan04}. We notice that going from Np to Pu the number of electrons in the 5$f_{7/2}$ sub-shell remains approximately the same, implying that by going from Np to Pu, we are filling with 1 electron more the 5$f_{5/2}$ sub-shell.

\subsection{C. Element specific magnetization curve recorded by XMCD}

\begin{figure}
\includegraphics[width=8.5cm]{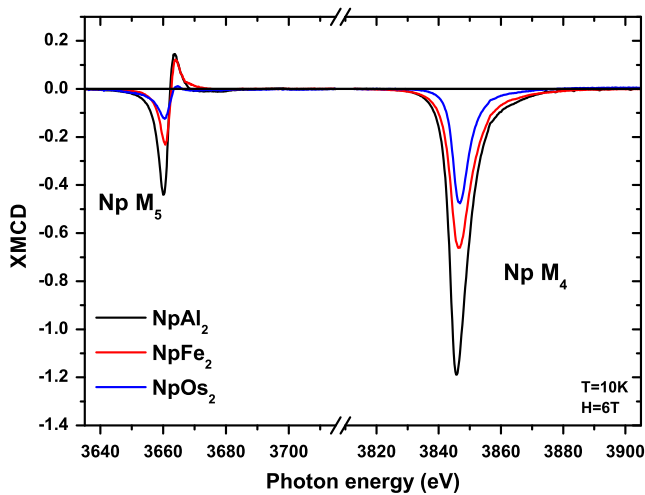}
\caption{(Color online) XMCD spectra measured at the Np $M_{5}$ and $M_{4}$ absorption edges for the polished NpX$_{2}$ samples in an applied field of 6T at 10K. The spectra have been corrected for self absorption effects and for the incomplete circular polarization rate.
\label{XMCDNp}}
\end{figure}

In Fig. \ref{XMCDNp} are shown the XMCD signals recorded at the Np $M_{4,5}$ absorption edges under magnetic field of 6T and at 10K in NpX$_{2}$ compounds. We observe that the XMCD signal at the Np $M_{4}$ edge is large and that its spectral shape consists of a single nearly symmetric negative peak that has no distinct structure. This dichroic signal at the $M_{4}$ edge is also characteristic for all uranium systems. The dichroic signal at the $M_{5}$ edge is nearly three times smaller than at the $M_{4}$. The spectral shape of the XMCD signal has an asymmetric $s$ shape with two peaks - a negative and a positive peak. However this positive peak is most pronounced in NpFe$_{2}$ and quasi absent for NpOs$_{2}$, as found in NpNiGa$_{5}$ \cite{okane09}. This asymmetric spectral shape, like for the $M_{5}$-edge of U, depends strongly on the hybridization, Coulomb, exchange, and crystal-field interactions. We will see in section IV that within a band structure model, this asymmetry can be well explained \cite{kunes01a}.

\begin{figure}
\includegraphics[width=8.5cm]{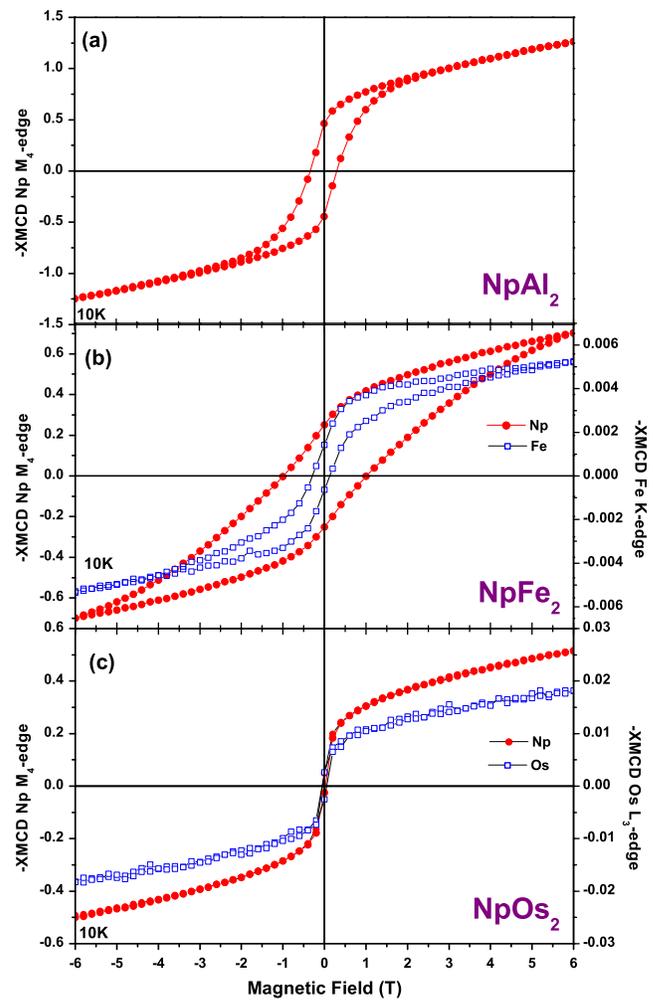}
\caption{(Color online) Element specific XMCD magnetization recorded at the maximum XMCD signal at the $M_{4}$ absorption edge (E=3846eV) measured at 10K for the polished (a) NpAl$_{2}$,(b) NpFe$_{2}$ and (c) NpOs$_{2}$ samples. Together is also shown the element specific XMCD magnetization recorded at the maximum XMCD signal at the Fe $K$ absorption edge (E=7113.7eV) and at the Os $L_{3}$ absorption edge (E=10877eV) for the polished NpFe$_{2}$ and NpOs$_{2}$ samples respectively.
\label{MXMCDNp}}
\end{figure}

Since one of the principal aims of the present investigation is to compare the moment determined by SQUID magnetization (Fig. \ref{MagnetizationNp}) with that deduced by the XMCD technique, we show in Fig. \ref{MXMCDNp} the XMCD signal at the $M_{4}$ absorption edge as a function of applied magnetic field from polished samples of the NpX$_{2}$ systems.

It is clear in comparing Fig. \ref{MXMCDNp} with Fig. \ref{MagnetizationNp} that the element specific magnetization curve recorded at the maximum XMCD signal at the $M_{4}$ absorption edge has a quite different form, and shows no sign of saturation for all the three NpAl$_{2}$, NpFe$_{2}$ and NpOs$_{2}$ polished samples. Surprisingly, one observes opening of the hysteresis loops. It seems, therefore, that magnetic domains at the sample surface (probing depth at the $M_{4}$ edge resonance is in the range of $\sim$200nm) are strongly constrained, and do not rotate with the applied field. We have then recorded the element-specific magnetization curve at the maximum XMCD signal recorded at Fe $K$ edge and Os $L_{3}$ edge. Since these signals are taken with photons of higher energy, the probing depth is larger, especially at the Os $L_{3}$ edge. They are clearly closer to saturation, as would be expected, as they penetrate further (at least one micron). It also interesting to notice that despite the unexpected shape of the hysteresis curves in Fig. \ref{MXMCDNp}, they do confirm one of the points made in the earlier studies. The magnetic anisotropy of NpFe$_{2}$ is very large, that of NpAl$_{2}$ smaller, and that of NpOs$_{2}$ almost negligible. This is consistent with the strong coercivity shown by the expanded hysteresis loop in NpFe$_{2}$ and its absence in the case of NpOs$_{2}$. Further, the coercive field of the hysteresis curve recorded at the Fe $K$-edge, as well as at the Os $L_{3}$-edge, is smaller than the one recorded at the Np $M_{4}$-edge. It means that the strong anisotropy of the magnetization at the surface, most probably caused by the polishing, clearly penetrates into the sample, despite the coercive field being smaller. As a direct consequence, we cannot extract meaningful parameters (such as the spin, orbital magnetic moments, and $\langle T_{z}\rangle$) from these data, and make a comparison with macroscopic measurements. Because of the very high anisotropies known to exist for NpFe$_{2}$ and, to a lesser extent, NpAl$_{2}$, (see Fig. \ref{MXMCDNp}), we judged it difficult to remove all the surface effects in these two materials, and concentrated on NpOs$_{2}$.

\begin{figure}
\includegraphics[width=8.5cm]{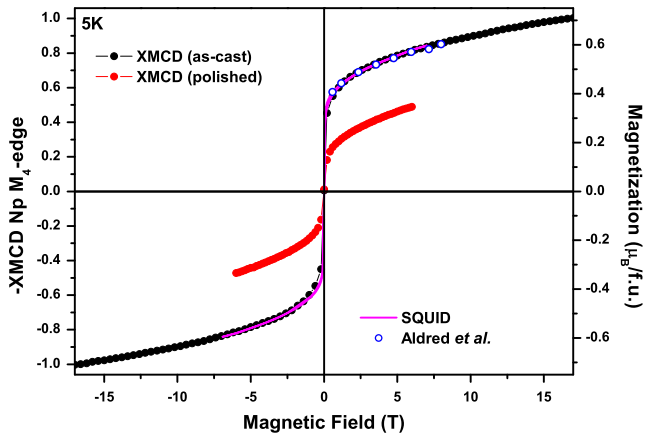}
\caption{(Color online) Element specific XMCD magnetization recorded at the maximum XMCD signal at the Np $M_{4}$ absorption edge (E=3846eV) measured at 5K for the as-cast and polished NpOs$_{2}$ samples. The earlier results from Aldred \textit{et al.} \cite{aldred76} are from magnetization experiments.
\label{MXMCDNpOs}}
\end{figure}

Let us now return to the element-specific magnetic characterization by XMCD.  As it can be seen from Fig. \ref{MXMCDNpOs}, the amplitude of the XMCD signal as a function of the magnetic field recorded at the $M_{4}$-edge is weaker for the polished sample of NpOs$_{2}$ than for the as-cast sample. Thus, the polished surface affects not only the approach to saturation, but also the absolute value of the signal. If we superimpose now these element-specific magnetization curves for the as-cast sample with the SQUID measurements, excellent agreement is obtained over the whole range of applied field.

Regarding the sample of PuFe$_{2}$, which was as-cast, (it is almost a single crystal since the sample came from the same boule as used for the single crystal studied by Wulff \textit{et al.} in Ref. \cite{wulff88}), the element specific magnetization curve recorded at the Pu $M_{4}$-edge shows saturation and similar hysteresis as the macroscopic measurements.

The conclusion from this XANES and XMCD characterization is that, for actinides systems, great care has to be taken to assure that surface effects do not play a role in comparing XMCD and magnetization results, especially of highly anisotropic materials.

\subsection{D. X-ray magnetic circular dichroism at An $M_{4,5}$ absorption edges}

The orbital contribution to the 5$f$ magnetic moment \cite{thole92} is obtained from the dichroic signal integrated over the pair of spin-orbit split excitations, $\Delta I_{M_{5}}+\Delta I_{M_{4}}$ normalized to the isotropic x-ray-absorption spectrum,

\begin{equation}\label{om}
\langle L_{z}\rangle = \frac{2 n_{h}}{I_{M_{5}}+I_{M_{4}}} (\Delta I_{M_{5}}+\Delta I_{M_{4}})
\end{equation}

A second sum rule \cite{carra92,carra93} correlates a linear combination of the partial dichroic signals $\Delta I_{M_{5}}$ and $\Delta I_{M_{4}}$ with the effective spin polarization $\langle S^{eff}\rangle$, which, in turn, is related to the spin operator through Eq. \ref{seff}, and given by

\begin{equation}\label{sm}
\langle S_{z}\rangle + 3\langle T_{z}\rangle = \frac{n_{h}}{I_{M_{5}}+I_{M_{4}}} (\Delta I_{M_{5}}-\frac{3}{2}\Delta I_{M_{4}})
\end{equation}

where $T_{z}$ is the component along the quantization axis of the magnetic-dipole operator. This operator, correlating the spin and position of individual electrons, is associated with the asphericity of the electronic cloud, distorted by crystal-field or spin-orbit effects, and with the spin anisotropy. $\langle T_{z}\rangle$ is therefore correlated to the charge and magnetic anisotropy \cite{collins95}.

The orbital and spin components of the total magnetic moment $\mu = -(\langle L_{z} \rangle + 2\langle S_{z} \rangle)~\mu_{B}$ can then be obtained from  XMCD spectra, together with an estimate of $\langle T_{z} \rangle$, if the value of the total moment $\mu$ and the occupation number of the $5f$ shell are known. Note that the magnetization data will give the total moment per formula unit, $\mu_{tot} = \mu_{An} + 2\mu_{X}$,  where $\mu_{X}$ is the contribution from the moments at the $X$ sites, and $\mu_{An} = \mu_{An}^{5f} + \mu_{cond}$ where $\mu_{cond}$, is the contribution, usually small, from the 6$d$7$s$ conduction band. There have been several attempts to study $\mu_{cond}$ combining magnetization and neutron data \cite{freeman76} and magnetic Compton scattering with XMCD \cite{kernavanois01}, and a rough estimate is that it is about $-10\%$ of the total moment. The reason for the negative sign is that the conduction electrons are polarized by the spin contribution at the actinide site and, as is well known, this spin contribution is antiparallel to the total moment in the light actinides, as the moment is dominated by the large orbital moment \cite{aldred75a}. Thus for NpOs$_{2}$, in making the comparison of the XMCD results with those from the magnetization, we shall assume that $\mu_{cond}$ is -0.1 $\mu_{Np}$, so that $\mu_{Np}^{5f} = 1.1 \mu_{Np}$ will be compared to the signal deduced from XMCD. In the case of PuFe$_{2}$ we use the 5$f$ moment deduced from the polarized-neutron study of a single crystal \cite{wulff88}.

As discussed earlier, the total signal as deduced from XMCD is unreliable for NpAl$_{2}$ and NpFe$_{2}$ since there is no saturation of the signal (see Fig. \ref{MXMCDNp}), so the determination of spin and orbital moments as well as the $T_{z}$ contribution is restricted to NpOs$_{2}$ and PuFe$_{2}$ samples.

\begin{table}
\caption{Experimental results for the Np and Pu 5$f$orbital and spin magnetic moments deduced from XMCD for NpOs$_{2}$ (scaled to 7T) and PuFe$_{2}$ as well as the $m_{md}$=-6$\langle T_{z}\rangle$$\mu_{B}$ contribution deduced combining results extracted from XMCD and SQUID or polarized neutrons measurements compared together with theoretical expectation values. Note that here we compare the measured values to model atomic calculations. These values will be compared to DFT claculations in Sec. IV.}
\label{tableExp}
\begin{ruledtabular}
\begin{tabular}{ccc}
  &  NpOs$_{2}$  &  PuFe$_{2}$ \\
\hline\\
$\mu_{L}^{5f-An}$ ($\mu_{B}$/An atom)           & 1.09(5)         &  1.98(5)      \\
$\mu_{S^{eff}}^{5f-An}$ ($\mu_{B}$/An atom)     & $-1.15(5)$      &  $-1.96(5)$   \\
$\mu_{L}^{5f-An}$/$\mu_{S^{eff}}^{5f-An}$       & $-0.95(8)$      &  $-1.01(8)$   \\
$\mu_{tot}$ ($\mu_{B}$/f.u.)                    & 0.60(1)         &               \\
$\mu_{tot}^{5f-An}$ ($\mu_{B}$/An atom)         & 0.46(5)         &  0.39(5)      \\
$\mu_{S}^{5f-An}$ ($\mu_{B}$/An atom)           & $-0.63(7)$      &  $-1.59(7)$   \\
\hline\\
$\mu_{L}^{5f-An}$/$\mu_{S}^{5f-An}$             & $-1.73(10)$     &  $-1.25(10)$  \\
Theory IC                                       & $-1.90$         &  $-1.35$      \\
\hline\\
$m_{md}$ ($\mu_{B}$/An atom)                    & $-0.52(3)$      &  $-0.37(3)$   \\
\hline\\
$m_{md}$/$\mu_{S}^{5f-An}$                      & $0.83(5)$       &  $+0.23(5)$   \\
Theory $LS$                                     & $-0.318$        &  $-0.693$     \\
Theory IC                                       & $0.555$         &  $-0.218$     \\
Theory $jj$                                     & $4.80$          &  $4.80$       \\
\end{tabular}
\end{ruledtabular}
\end{table}

\begin{figure}
\includegraphics[width=8.5cm]{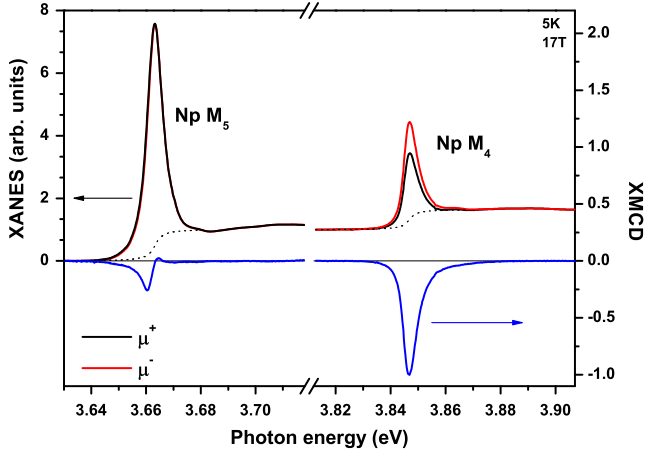}
\caption{(Color online) XANES measured with right and left circularly polarized X-rays and associated XMCD spectrum recorded at the Np $M_{4,5}$ absorption edges under 17T and at 5K for the as-cast NpOs$_{2}$ sample. The spectra have been corrected for self-absorption effects and for incomplete circular polarization rate.
\label{SampleNpOs}}
\end{figure}

\begin{figure}
\includegraphics[width=8.5cm]{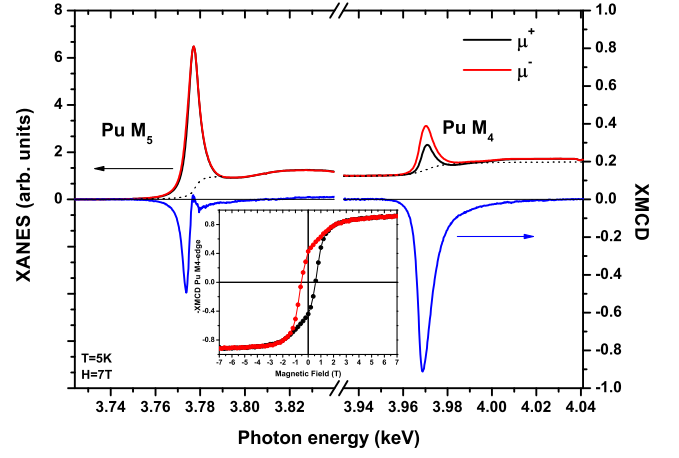}
\caption{(Color online) XANES measured with right and left circularly polarized X-rays and associated XMCD spectrum recorded at the Pu $M_{4,5}$ absorption edges under 17T and at 5K for the as-cast PuFe$_{2}$ sample. The spectra have been corrected for self-absorption effects and for incomplete circular polarization rate. In the insert is shown the element specific XMCD magnetization recorded at the maximum XMCD signal at the Pu $M_{4}$ absorption edge (E=3846eV) measured at 5K.
\label{SamplePuFe}}
\end{figure}

The XANES and XMCD signal recorded at the $M_{4,5}$ edges for the 2 compounds discussed in this section are shown in Figs. \ref{SampleNpOs} and \ref{SamplePuFe}.

We have first determined the 5$f$ orbital magnetic contribution for the NpOs$_{2}$ and PuFe$_{2}$ samples. The results are given in the Table \ref{tableExp}. As for uranium compounds, the 5$f$ orbital contribution is large and parallel to the external magnetic field. The difficulty in determining the 5$f$ spin magnetic moments, as already mentioned above, lies in the fact that the magnetic dipole contribution cannot be measured directly. The spin magnetic moment can be individually determined either by taking the $\langle T_{z} \rangle$ contribution from DFT or multiplet calculation (for a free-ion without crystal-field interaction) or by combining results from neutron diffraction and/or macroscopic measurements. Concentrating on NpOs$_{2}$, for which we have XMCD data at both the Np $M$ and Os $L$ edges, the XMCD results, assuming $n_{h}=10$,  give a 5$f$-orbital magnetic moment $\mu_{L}^{5f-Np}=1.09\mu_{B}/Np$ from Eq. (3). With a total magnetic moment $\mu^{5f-Np}=0.46\mu_{B}$ (taking into account the Os magnetic contribution extracted from XMCD, see section III.E, and the $spd$ estimated contributions of Np), we obtain a spin magnetic moment on Np sites $\mu_{S}^{5f-Np}= -0.63\mu_{B}/Np$. It is noteworthy that $\mu_{L}^{5f-Np}/\mu_{S}^{5f-Np}= -1.73$ is close to the Np$^{3+}$ free-ion value ($-1.90$). From Eq. \ref{sm}, we can then extract $\langle T_{z} \rangle$, which is $+0.09$. In fact, it is better to consider the magnetic dipole contribution $m_{md}$ normalized by the spin moment $\mu_{S}^{5f}$ on the Np atom, and this can be compared directly to calculated values in Table \ref{tableExp}. $m_{md}$/$\mu_{S}^{5f}$=+0.83 whereas in Np$_{2}$Co$_{17}$ this value is +1.36. The closest theoretical value is that of the intermediate coupling of +0.56. A similar analysis can be done for PuFe$_{2}$, but instead of using macroscopic measurements, since we do not have access to the Fe magnetic moments, we used the results of Ref. \cite{wulff88} using neutron diffraction. From the XMCD results, assuming $n_{h}=9$, one obtains a 5$f$-orbital magnetic moment $\mu_{L}^{5f-Pu}=1.98\mu_{B}/Pu$ from Eq. \ref{om}. The deduced spin moment on Pu sites $\mu_{S}^{5f-Pu}=-1.59\mu_{B}/Pu$, and the ratio $\mu_{L}^{5f-Np}/\mu_{S}^{5f-Np}=-1.25$. Then for the Pu $m_{md}$/$\mu_{S}^{5f}$=+0.37, which is small and closest again to the IC value of $-0.22$.

The neutron result for PuFe$_{2}$ for $\mu_{L}^{5f-Pu}/\mu_{S}^{5f-Pu}$ is $-1.20 \pm 0.05$, in excellent agreement with the XMCD result of $-1.25 \pm 0.05$. These values are slightly below the IC theory value of -1.35 indicating a hybridization with the Fe 3$d$ electrons, as already discussed in Ref. \cite{wulff88}. Moreover, the individual values for the orbital and spin moments determined by neutron diffraction are $\mu_{L}^{5f-Pu}=2.3 \pm 0.3 \mu_{B}$ and $\mu_{S}^{5f-Pu}=-1.9 \pm 0.3 \mu_{B}$, which are in reasonable agreement with the values determined from the XMCD in the table. We expect the neutron results to be reliable for the ratio $\mu_{L}^{5f-An}/\mu_{S}^{5f-An}$ but not so accurate for the individual values. The XMCD technique has better precision, but shows that the neutron values are reliable.

The 5$f$ moment at the Np site in NpOs$_{2}$ is $0.46 \pm 0.03 \mu_{B}$ in good agreement with the value of $0.40 \pm 0.04 \mu_{B}$ deduced from the hyperfine field measured in the M\"{o}ssbauer experiments of Aldred \textit{et al.} \cite{aldred76}.

From the XMCD measurements, we can conclude that, similarly to uranium compounds, the largest contribution to the 5$f$ magnetic moment arises from the orbital contribution, which is parallel to the external field. The spin contribution is antiparallel to the orbital one. In both NpOs$_{2}$ and PuFe$_{2}$, the expectation value of the magnetic-dipole contribution is small and is compatible with the value predicted for 5$f$ electrons calculated in intermediate-coupling (IC) approximations.

NpOs$_{2}$ has one of the smallest ordered moments of any Np-intermetallic compound, and this value has been difficult for theory to reproduce.

\subsection{E. X-ray magnetic circular dichroism at Os $L_{2,3}$ absorption edges}

\begin{figure}
\includegraphics[width=8.5cm]{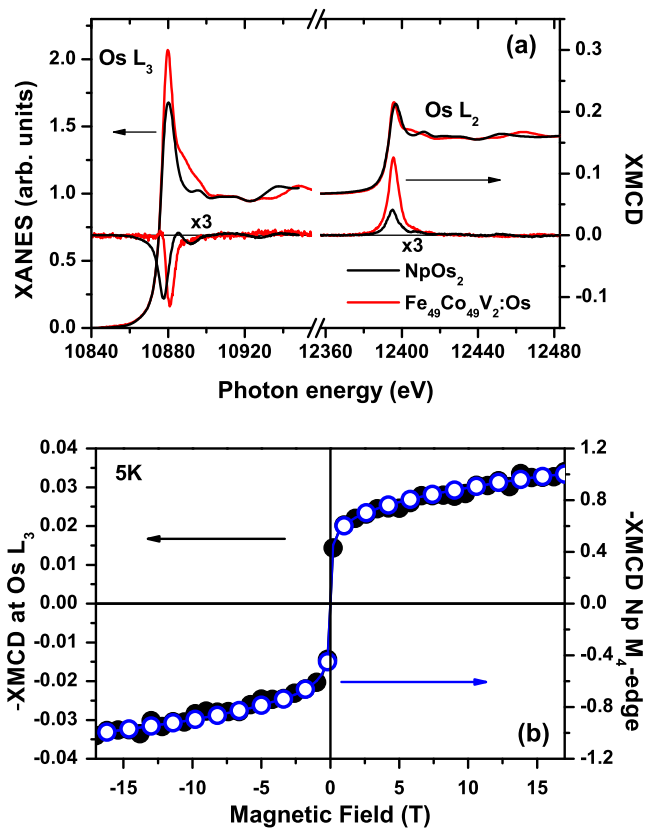}
\caption{(Color online) (a) Isotropic XANES and XMCD spectra recorded at the Os $L_{2,3}$ absorption edges under 17T and at 5K for the as-cast NpOs$_{2}$ sample and under 6T and 300K for the Co$_{49}$Fe$_{49}$V$_{2}$ sample doped with 3\% Os. The XMCD spectra recorded on the as-cast NpOs$_{2}$ have been multiplied by a factor 3. In (b) is shown the element specific XMCD magnetization recorded at the maximum XMCD signal at the Os $L_{3}$ absorption edge (E=10877eV) (solid points) measured at 5K together with the element specific XMCD magnetization recorded at the maximum XMCD signal at the Np $M_{4}$ absorption edge (E=3846eV) (open points) for the as-cast NpOs$_{2}$ sample.
\label{SampleOs}}
\end{figure}

Figure \ref{SampleOs} shows the XANES and XMCD signals recorded at the $L_{2,3}$ absorption edges of Os for the as-cast NpOs$_{2}$ polycrystalline sample, after corrections of self-absorption effects, the latter being small.

Using a similar analysis procedure as described in Ref. \cite{wilhelm01}, the resulting application of the sum rules at the $L_{2,3}$ absorption edges of Os, using a number of 5$d$ holes of 3.07%
\footnote{This was deduced from the changes in white lines intensities in NpOs$_2$ with respect to Os impurities in the permendur reference sample, in which the estimated number of 5$d$ holes is n$_{h}^{5d}$=3.70.} and neglecting the $\langle T_{z}\rangle$ contribution, gives for the Os 5$d$ states a spin and orbital induced magnetic moment contributions under a field of 17T and 5K. We then re-scale the values to 7T applied field using the element selective magnetization curve shown in Fig.\ref{SampleOs}b and obtain:

$\mu_{L}^{5d-Os} = + 0.013 \pm 0.005 \mu_{B}/Os$\\
$\mu_{S}^{5d-Os} = + 0.080 \pm 0.005 \mu_{B}/Os$\\
$\mu_{tot}^{5d-Os} = + 0.093 \pm 0.01 \mu_{B}/Os$\\

 The Os 5$d$ contribution to the total magnetization is $0.18/0.60 \sim 30\%$. The total moment (neglecting the \textit{sp} contributions) is parallel to the Np moment. The induced spin contribution dominates over the orbital contribution, however, it is noteworthy that the induced orbital and spin magnetic moments are aligned parallel. This is different from Os impurities in Permendur alloy (Fe$_{49}$Co$_{49}$V$_2$), where the induced spin and orbital magnetic moments are aligned antiparallel. Theoretically, Tyer \textit{et al.}, \cite{tyer03}, show that for 5\% Os (which has $\sim$4 5$d$ holes in the atomic state) diluted in a Fe matrix, the spin moment should be positive, but the orbital moment should be negative, as we have found experimentally.

\section{IV. Electronic structure calculations}

We have performed electronic structure calculations of the NpX$_{2}$ (X=Al, Fe, Os) and PuFe$_{2}$ compounds using the WIEN2k \cite{wien2k} package, which implements the full-potential linearized augmented plane waves method with local orbitals to solve the Kohn-Sham equations. To deal with the strongly correlated nature of unfilled $f$-electron shells, we have performed both local spin density approximation (LSDA) calculations as well as LSDA+$U$ calculations, which include explicitly the strong onsite Coulomb interactions in the 5$f$-electron subsystem. We have tested both the most common variants of the LSDA+$U$, namely the around mean-field (AMF \cite{ldauamf}) and fully localized limit (FLL \cite{fll}) formulations. The spin orbital (SO) interaction, which is essential for actinide atoms, was included in the second variational step \cite{kunesso}, where also the orbital potential was introduced, including its spin cross-term. In some of the calculations performed for NpOs$_{2}$, see below, the SO interaction was explicitly switched off on selected atoms, to probe its influence on the spin and orbital moments of the two elements. The structure parameters were set to 7.189{\AA}, 7.528{\AA}, 7.144{\AA}, and 7.785{\AA} for PuFe$_{2}$, NpOs$_{2}$, NpFe$_{2}$, and NpAl$_{2}$, respectively. The atomic sphere radii were set to 2.9 a.u., 2.5 a.u., 2.35 a.u., and 2.5 a.u., for actinide and Os, Fe, and Al, respectively. The $RK_\text{max}$ parameter was set to 8.0, which leads to a basis size of approximately 500-550, depending on lattice parameter and muffin-tin radii. In total, 10000 $\mathbf{k}$-points were used, leading to 726 $\mathbf{k}$-points in the irreducible wedge of the Brillouin zone. These parameters were checked to provide well-converged electronic structure calculations.

\begin{figure}
 \includegraphics[width=5.8cm,angle=270]{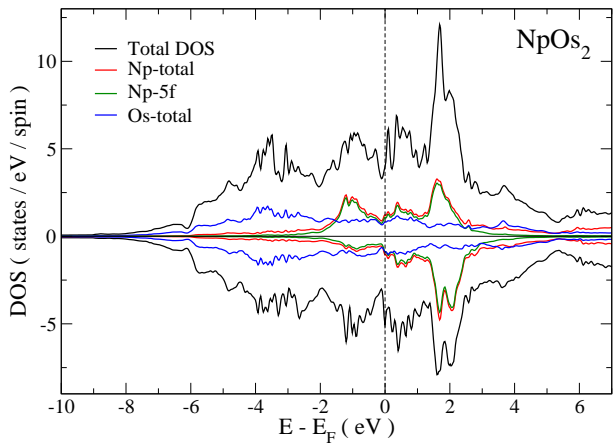}
 \caption{(Color online) Spin- and atom-resolved density of states calculated for NpOs$_2$ with the AMF variant of LSDA+$U$ and $U=1.0$eV.
 \label{NpOs2DOS}}
\end{figure}

\begin{figure}
 \includegraphics[width=5.8cm,angle=270]{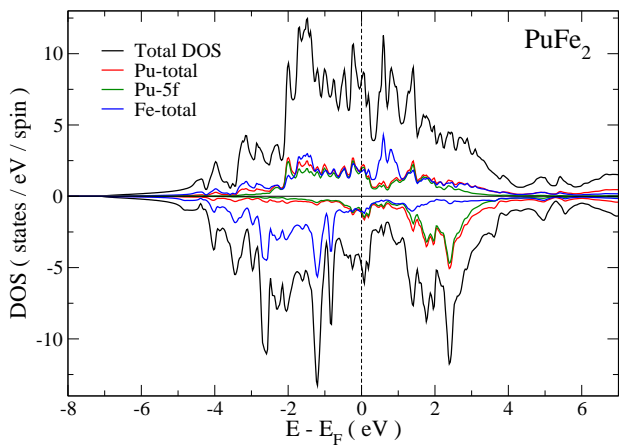}
 \caption{(Color online) Spin- and atom-resolved density of states calculated for PuFe$_2$ with the FLL variant of LSDA+$U$ and $U=1.0$eV.
 \label{PuFe2DOS}}
\end{figure}

In Figs.~\ref{NpOs2DOS} and ~\ref{PuFe2DOS}, we show the density of states (DOS) of NpOs$_{2}$ and PuFe$_{2}$, respectively, for values of $U$ and variants LSDA+$U$; which provide the best agreement with experimental XMCD spectra, see below. One can observe large exchange splitting in the Np 5$f$ states, which is mainly responsible for the different spin-up and spin-down projection of the total DOS. The two spin projected DOSes for Os atoms look very similar to each other reflecting the small induced spin moment on this atom in NpOs$_2$, see Table~\ref{tableTheory}. The unoccupied states are dominated by sharp peaks of Np-5$f$ character. It is transitions to these states that dominates the x-ray absorption spectra. A similar situation for unoccupied states applies also for the case of PuFe$_2$, see Fig.~\ref{PuFe2DOS}. The strongest feature in the unoccupied states is formed by a 2eV wide Pu-5$f$ band starting approximately 2eV above the Fermi level. For the occupied states it is the exchange-split Fe-3$d$ band that represents a dominant contribution.

\begin{figure}
\includegraphics[width=8.6cm]{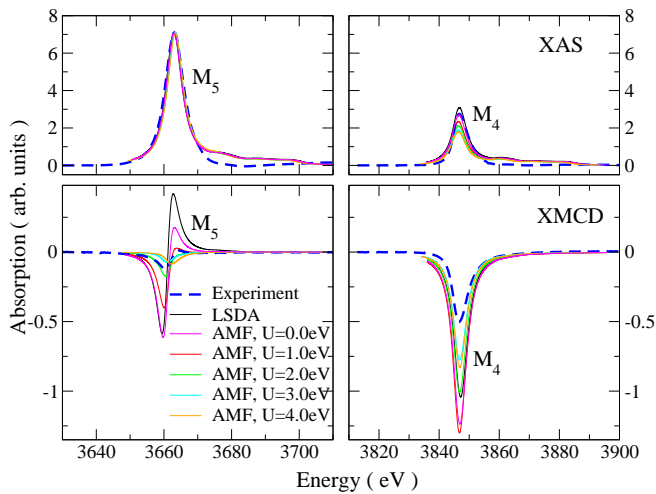}
\caption{(Color online) Theoretical and experimental XAS and XMCD of $M_{4,5}$ absorption edges of NpOs$_{2}$. Theoretical results are presented for various values of the $U$ parameter in the LSDA+$U$ potential, as indicated in the legend.
\label{NpOs2Theory}}
\end{figure}

The XAS have been simulated by the initial state approximation in which the absorption strength at the $M_{4,5}$ edges is a function of the projected density of states of unoccupied 5$f$ states of the actinide element. The energy-dependent dipole transition matrix elements have been included. For a description of the implementation in the FLAPW basis we refer to Refs. \cite{kunes01a,kunes01b}. Figure \ref{NpOs2Theory} summarizes the resulting XAS spectra of NpOs$_{2}$ and their difference - the XMCD. For every system treated we have performed several independent simulations - varying the strength of the onsite Coulomb repulsion characterized by the $U$ parameter in the LSDA+$U$ method. Typically we have calculated the spectra for $U$ within the range from 0 to 4 eV. The value of $J$ was set to 0.6 eV for both Np and Pu 5$f$ electrons.

Due to of the strong influence of the spin-orbit interaction on the 3$d$ core shell, the 3$d_{3/2}$ and 3$d_{5/2}$ core levels are split by approximately 200 eV. To calculate both edges within the same energy window would thus require evaluation of the unoccupied states to more than 200 eV beyond the Fermi level, which is not feasible in electronic structure codes using linearized basis sets. On the other hand, it is reasonable to assume that the unoccupied 5$f$ states will be almost entirely present within a few eV above the Fermi level (as is supported by DOS curves, Figs.~\ref{NpOs2DOS} and ~\ref{PuFe2DOS}) and that transitions to dipole-allowed continuum states would form a very wide feature-less contribution that would not have any visible influence on the spectra. A similar argument applies for the unoccupied $p$-states well above the Fermi level. Therefore, we have calculated each edge separately in its energy range, considering states up to approximately 50 eV above the Fermi level. A sudden monotonous decrease in the calculated intensity approximately 50 eV above the edge onset is artificial and originates from this cut-off. Also a certain inaccuracy can be expected at energies well beyond approximately 20 eV above the Fermi level due to the linearization errors present in the basis functions, something which should be kept in mind, when interpreting the calculated spectral features far above the edge onset.

All spectra were broadened with a Lorentzian of 2.5eV half-width at half-maximum, which provided good match with experimental widths of absorption peaks. This broadening can be interpreted as a combined effect of instrumental broadening and life-time broadening, respectively. The intensity of the $M_{5}$ XAS edge was normalized to the same peak value as the experimental one. The same scaling factors were applied also for $M_4$ edges and XMCD spectra, in order to preserve their relative magnitudes.

Comparing the theoretical and experimental XAS spectra of NpOs$_{2}$, the first-principles calculations reproduce the experimental findings with good qualitative accuracy with regard to the peak shapes, distance and branching ratio. These features are only very weakly sensitive to the chosen model of treatment of the 5$f$ electrons. A certain degree of sensitivity on the strength of the onsite Coulomb correlations $U$ can be seen on the intensity of the $M_{4}$ peak. Best agreement between theory and experiment is obtained for small values of $U$.

The XMCD spectra reveal much more details. Generally, the XMCD signal on the $M_{5}$ edge is very weak, compared to the XAS intensity. In the experimental spectra there has an $s$-shape, starting with negative values, switching the sign to a positive peak and monotonously decreasing to zero.  The origin of such an XMCD shape has been outlined previously \cite{kunes01a}.

Considering the sum rules expression (Eqs. \ref{om} and \ref{sm}), the sine-like shape implies that the energy-integral of the XMCD spectrum over the $M_{5}$ edge will yield a very low value and, hence, it is mainly the $M_{4}$ edge signal that will determine the local spin and orbital moment magnitudes. Indeed, the XMCD signal at the $M_{4}$ edge is substantially larger, showing only a simple negative peak of rather large relative magnitude compared to the XAS $M_{4}$ peak. This indicates that for the three Np-based compounds, the ratio of the orbital and spin magnetic moments will be approximately the same if $\langle T_{z} \rangle$ is not changed. This finding should remain valid also for NpAl$_{2}$ and NpFe$_{2}$ despite that in experiments we have not reached magnetic saturation.

The theoretical XMCD spectra at the $M_{5}$ edge show an interesting, rather strong sensitivity on the chosen model of the description of the 5$f$ states. The small $s$-like signal is best reproduced by low $U$ values. Even the LSDA reproduces the sine-like XMCD signal, however, it overestimates its magnitude. The sensitivity of the XMCD spectrum shape on the chosen 5$f$ model is much reduced at the $M_{4}$ edge, although the magnitude changes visibly, when adding the onsite Coulomb correlation effects into the calculation. Our first-principles calculations thus reveal a strong sensitivity of the computed XMCD spectra to the degree of 5$f$ electron localization, as described by the Coulomb $U$. In comparison with an experimental XMCD spectrum, this feature provides a means of establishing the appropriate Hubbard $U$ value for a specific actinide compound. Here we find that, generally, for the NpX$_{2}$ compounds, they appear to be best described by low $U$ values ($U$= 0 -1 eV), i.e. \textit{the 5$f$ electrons behave mostly as itinerant}.

For the NpOs$_{2}$ compound we have performed a more detailed study of the electronic structure, in order to compare to experimental findings, including the Os local magnetic properties, as extracted from the experiment. Namely, we have focused on the influence of the spin-orbital interaction of both Np and Os atoms on their spin and orbital magnetic moments. The WIEN2k package offers the possibility to selectively turn the spin-orbital interaction on and off within the atomic spheres of individual atoms. Using this feature, we have calculated the electronic structure properties of NpOs$_{2}$ in four configurations: 1) full spin-orbital calculation, 2) no spin-orbital interaction, and 3) and 4) spin-orbital interaction present only on the Np or on the Os atom, respectively.

\begin{table*}
\caption{Summary of the magnetic characteristics of NpOs$_{2}$ based on electronic structure calculations. The spin, orbital and total magnetic moments ($\mu_S$, $\mu_L$, $\mu_{tot}$) and magnetic dipole term ($T_z$), all in Bohr magnetons, are shown as a function of Coulomb $U$ parameter and variant of the LSDA+$U$ orbital functional, for both Np and Os atoms, respectively. The exchange $J$ parameter is fixed at 0.6eV in all calculations (of course, except for the LSDA ones). We chose a sign convention in which the total moments per formula unit ($\mu_{tot}^{5f\text{-Np}}$ + 2$\mu_{S}^{5d\text{-Os}}$)  are always positive. Recall from Table~\ref{tableExp} that the experimental values for Np 5$f$ states at 7T are $\mu_{L}$ = +1.09(5); $\mu_{S}$ = -0.63(7);  giving $\mu_\text{tot}$ = +0.46(5),  $\langle T_{z}\rangle$ = +0.09, and for Os 5$d$ states (assuming $\langle T_{z}\rangle$ = 0) are $\mu_{L}$ = +0.016(5) and $\mu_{S}$ = +0.096(5) giving $\mu_\text{tot}$ = 0.11(1).}
\label{tableTheory}
\begin{ruledtabular}
\begin{tabular}{cccccccc}
  &  $\mu_{S}^{5f\text{-Np}}$  &  $\mu_{L}^{5f\text{-Np}}$  &  $\mu_\text{tot}^{5f\text{-Np}}$  &  $\langle T_{z}\rangle ^{5f\text{-Np}}$  &  $\mu_{S}^{5d\text{-Os}}$  &  $\mu_{L}^{5d\text{-Os}}$  &  $\mu_{tot}^{5d\text{-Os}}$ \\
\hline\\
 LSDA	        & -2.267 & 2.232 & -0.035  & 0.024  & 0.120  &  0.029 &  0.149  \\
 AMF 0.0eV	& -2.049 & 2.957 &  0.908  & 0.067  & 0.108  &  0.035 &  0.143  \\
 AMF 1.0eV	& -1.616 & 3.194 &  1.578  & 0.136  & 0.083  &  0.031 &  0.114  \\
 AMF 2.0eV	& -1.030 & 2.793 &  1.763  & 0.198  & 0.047  &  0.010 &  0.057  \\
 AMF 3.0eV	& -0.626 & 2.081 &  1.455  & 0.181  & 0.024  & -0.007 &  0.017  \\
 AMF 4.0eV	& -0.363 & 1.526 &  1.163  & 0.150  & 0.006  & -0.018 & -0.012  \\
 FLL 0.0eV	& -1.429 & 2.324 &  0.895  & 0.080  & 0.073  &  0.026 &  0.099  \\
 FLL 1.0eV	& -1.959 & 3.537 &  1.578  & 0.113  & 0.105  &  0.039 &  0.144  \\
 FLL 2.0eV	& -1.972 & 3.788 &  1.816  & 0.140  & 0.104  &  0.035 &  0.139  \\
 FLL 3.0eV	& -1.902 & 3.893 &  1.991  & 0.169  & 0.092  &  0.028 &  0.120  \\
\end{tabular}
\end{ruledtabular}
\end{table*}

Generally, using the AMF variant of the LSDA+$U$ causes the spin moment to decrease and the orbital moment behaves non-monotonically. It grows quickly already for small $U$ and then decreases in magnitude. In the FLL variant of the LSDA+$U$, both spin and orbital moments are less sensitive to $U$, although there is a weak increase of the orbital moment magnitudes with growing $U$. The $\langle T_{z}\rangle$ term expectation value is substantially influenced by inclusion of the onsite Coulomb correlations, but the magnitude remains within the 0.1-0.2 range (on the Np atom).

The Os spin magnetization follows the evolution of the spin magnetic moment on the Np atom, which indicates that the Os magnetization is of induced nature, far from the atomic regime and Hund's rules. The orbital moment of Os is very small and its value (not magnitude) grows as a function of $U$, passing from negative to positive values in the case of the AMF variant of LSDA+$U$. The magnetic dipole term on Os is predicted to be of the order of 10$^{-3}$ $\mu_{B}$.

The occupation of the 5$f$-shell of Np remains approximately 3.65, independent of the value of $U$ and chosen variant of the double-counting term (FLL or AMF), although a weak decrease can be observed with increasing $U$ in the FLL variant. This occupation is close to an expected 5$f^{4}$ configuration (Np$^{3+}$), consistent with the interpretation of the shift of the Np M\"{o}ssbauer line\cite{aldred76}. The occupation number of the 5$d$ states of Os is also more-or-less insensitive to $U$ and variant of LSDA+$U$ potential - approximately 5.29 5$d$ electrons are obtained within the Os atomic sphere (\textit{i.e.}, $n_h^{Os}$=4.71).

Switching off the SO interaction on the Os atom does not seem to have a strong influence on the local charge distribution and magnetism of both Np and Os atoms. The most pronounced difference can be seen in the orbital magnetic moments of Np, especially in the FLL calculations or AMF calculations with larger $U$. The situation is different when the SO interaction is switched off on Np atoms (while it is kept on Os atoms). In this case, we observe strong changes in the local electronic structures of both atoms. The spin moment on Np becomes substantially larger, while the orbital moment becomes much more sensitive to the $U$ value. There is also a large effect on the magnetic dipole term of Np, which becomes substantially reduced. An interesting influence can be seen on the spin moment of the Os atoms, which become enhanced by switching off the SO interaction on Np atom. This somewhat unusual observation can be explained by hybridization effects between the Np and Os electrons, because the spin moment grows on Np as well, providing further support to the suggestion that the magnetism of Os is of induced nature. Switching off the SO interaction on both atoms almost entirely suppresses the orbital moments and magnetic dipole terms on both atoms. In DFT calculations \cite{tyer03} the orbital and spin moments of Os as impurities in Fe metal are antiparallel. In NpOs$_{2}$, they are parallel, as shown by both theory and experiment. This is a consequence of a complicated interplay of hybridization, Coulomb correlation. and relativistic effects, which are well captured by the present calculations.

\begin{figure}
\includegraphics[width=5.8cm,angle=270]{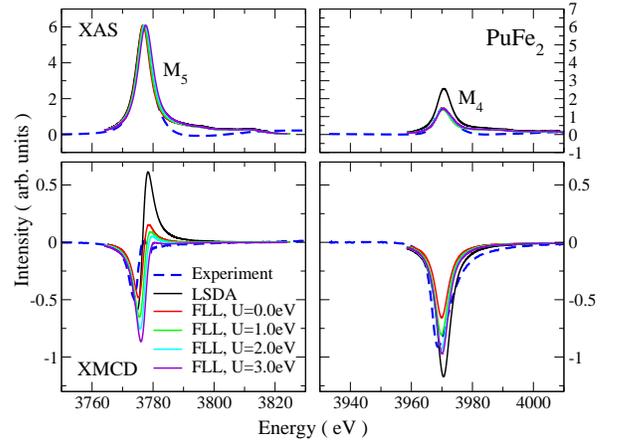}
\caption{(Color online) Theoretical XAS and XMCD of $M_{4,5}$ edges of PuFe$_{2}$. Calculations are presented for various values of the $U$ parameter in the LSDA+$U$ potential of FLL variant, as indicated in the legend.
\label{PuFe2Theory}}
\end{figure}

Comparing these results to previous calculations \cite{eriksson90b}, which used orbital polarization corrections (OPC) to the exchange-correlation potential, the individual spin and orbital moments on Np are of more realistic sizes in the present calculations. However, the OPC provides the total moment in better agreement with experiment than our LSDA+$U$ calculations. It is a well-known effect that OPC tends to overestimate the spin and orbital potentials and, in some cases, their cancellation fortuitously leads to a rather good agreement with experiment, as found for NpOs$_{2}$ \cite{eriksson90b}.

For PuFe$_2$, it is the FLL variant of LSDA+$U$ \cite{fll} that provides the most satisfactory agreement with experimental spectra, see Fig.~\ref{PuFe2Theory}. The AMF LSDA+$U$ calculations lead to a far too weak XMCD signal at the $M_5$ edge. While the FLL calculations were rather unstable for low values of $U$, at $U=1$eV we obtained a good representation of the experimental XAS and XMCD spectra. From the theory we obtained spin moment of $\mu_{S}^{5f\text{-Pu}}=-1.76\mu_B$ and orbital moment of $\mu_{L}^{5f\text{-Pu}}=2.58\mu_B$ on Pu atoms, leading to a total moment of $\mu_{tot}^{5f\text{-Pu}}=0.82\mu_B$. The spin moment of Fe atoms is of a very similar magnitude to the Pu spin moment, namely $\mu_{S}^{3d\text{-Fe}}= 1.73\mu_B$. The theoretical Pu moments somewhat overestimate the measured ones, see Table \ref{tableExp}, on the other hand the orbital to spin moment ratio is in a reasonable agreement with experiment.

\section{V. Discussion and Conclusions}

\subsection{A. Actinides atoms}

Earlier experimental work was done on these materials at ANL in the 1970s \cite{lam72,harvey74,aldred74,aldred75a,aldred75b,aldred76,lander77,aldred79}, and the last efforts in theory dates from 1990s \cite{eriksson89,eriksson90a,eriksson90b,eriksson90c,severin93,shishidou99,antonov03a}. In all cases, especially for PuFe$_{2}$, where neutron experiments were done on a single crystal, our experimental results are in good agreement with previously determined quantities. XMCD has not been used previously for these materials and allows a more precise determination of the orbital ($\mu_{L}$), spin ($\mu_{S}$) moments and the magnetic dipole operator ($\langle T_{z} \rangle$).

With experiments on NpOs$_{2}$ we have confirmed that this compound has an exceptionally small total moment of $0.46 \mu_{B}$ on the Np atoms. In addition, our measurements have shown that this small moment is not a result of a major cancellation of two large but opposing orbital and spin moment, but instead both of them are reduced. Note that DFT calculations predict for the AMF variant of LSDA+$U$ with $U=0$ a ratio of -1.44, and for $U=1$ a ratio of -1.98, respectively. The experimental value is between these two, in accord with our earlier statement that small value of $U$ seem to be suitable for Np-based Laves phases. The ratio $\mu_{L}^{5f-Np}/\mu_{S}^{5f-Np}= -1.73 \pm 0.10$, which is only slightly reduced from the theoretical value of $-1.90$ for IC. The reduced values of both the orbital and spin moments represent a particularly difficult challenge for theory.

The result for the branching ratios (Table \ref{tableBR}) gives values of the spin-orbit operator per hole [$\langle W^{110}\rangle$/$n_{h}$-$\Delta$] for the Np compounds of between -0.34 to -0.36, which is somewhat larger (by $\sim$ 0.06) than the theoretical atomic value of $-0.417$ for IC. In good agreement with experiment, our DFT calculations suggest values between -0.30 and -0.38 with $U$ between 0 and 1eV. Similarly, for PuFe$_{2}$ the value of this quantity is -0.51, that is close to the theoretical atomic value of -0.567 (DFT value is -0.50 for FLL with $U$=1.0eV). Since the $\Delta$ values are small, these differences could arise either from the number of holes (10 for Np$^{3+}$ and 9 for Pu$^{3+}$) being too large, or from a change in the spin-orbit interaction caused by hybridization. Thus, although we are close to intermediate coupling values throughout, there are differences, and these can be ascribed to the effects of the hybridization with other electron states. Moreover, by comparing the present compounds with Np$_{2}$Co$_{17}$ \cite{halevy12}, where the experimental value of this parameter is -0.392, such a conclusion is reinforced. The value found for Np$_{2}$Co$_{17}$ is much closer to the IC theory value than found in the present NpX$_{2}$ materials. In Np$_{2}$Co$_{17}$ the conclusions are that the Np 5$f$ states are not strongly hybridized with the Co 3$d$ states.

Rather than discuss $\langle T_{z} \rangle$ we have chosen to present the term $m_{md}$/$\mu_{S}^{5f}$ (recall that $m_{md} = -6 \langle T_{z} \rangle$, and $\mu_{S}^{5f} = -2 \langle S_{z} \rangle$) as this better represents the influence of the dipole operator, and can be more easily compared to theory. The term is expected to be larger for a lower symmetry, as in the hexagonal symmetry found in Np$_{2}$Co$_{17}$ \cite{halevy12}. The value found in NpOs$_{2}$ of +0.83 is slightly larger than the IC value (+0.56) but less than the value of +1.36 found in Np$_{2}$Co$_{17}$. On the basis of the values found in UFe$_{2}$ \cite{dalmasdereotier98} it was proposed that the expectation value of the magnetic dipole operator is smaller when there is strong hybridization and a degree of itinerancy of the actinide 5$f$ electrons. There might be some evidence for this, especially when comparing the values in NpOs$_{2}$, an itinerant system, with those of Np$_{2}$Co$_{17}$, a localized system, but we note that the expectation value of the dipole operator is far from zero in NpOs$_{2}$.

\subsection{B. Transition metal Os atoms}

One of the most interesting observations of the present study is that there is a relatively large moment induced on the Os atoms in NpOs$_{2}$. This was not considered in the earlier work, and indeed demonstrates the power of the XMCD technique to examine sites in a compound other than those expected to carry magnetic moment. The total Os 5$d$ moment (at an applied field of 7 T) is +0.09 $\mu_{B}$. Of this the spin moment dominates at +0.08 $\mu_{B}$, and the small orbital moment is parallel to the spin moment, in contrast to other investigations \cite{wilhelm01,tyer03}. Thus, although hybridization is clearly present in NpOs$_{2}$ the extent and influence of both the interatomic (Np)5$f$ - (Np)5$f$ and the (Np)5$f$ - (Os)5$d$ interactions have to be understood before a full description of this material can be obtained. In NpOs$_{2}$ the induced 5$d$ polarization contributes some 30\% of the total magnetism per formula unit, so it is by far from negligible.

A limited number of studies have been made on the induced magnetization on 5$d$ elements in ferromagnetic matrices. Wilhelm \textit{et al.} \cite{wilhelm01} showed that for the lighter elements of the 5$d$ series (specifically W) the induced spin and orbital moments on the ligand should be coupled antiparallel, whereas for the heavier elements (specifically Ir and Pt) the coupling should be parallel. Regarding the induced orbital magnetic contribution, especially in the case of W and Ir, this contribution may be either aligned parallel or antiparallel to its spin induced magnetic moment \cite{wilhelm01,schutz93,krishnamurthy06,lagunamarco10}. Those experimental observations of a sign reversal of the orbital magnetism in transition-metal-based itinerant magnetic system were initially reported by Wilhelm \textit{et al.} \cite{wilhelm01} for W in Fe/W multilayers and by Krishnamurthy \textit{et al.} \cite{krishnamurthy06} for Ir in Co$_{100-x}$Ir$_{x}$ alloys.

We compare in Fig. \ref{SampleOs}(a) the XMCD spectra at the Os $L_{2,3}$ edges for polycrystalline Permendur doped with 3\% of Os with those for NpOs$_{2}$. In the case of this alloy, $\mu_{S}^{5d-Os} = + 0.50 \pm 0.005 \mu_{B}/Os$ and the orbital contribution is $-0.045 \pm 0.005 \mu_{B}/Os$, confirming the Tyer \textit{et al.}, \cite{tyer03} remark that the orbital and spin contributions should be anti-parallel for Os impurities in a Fe matrix. In fact, reference to the sum rules (Eqs. \ref{om} \& \ref{sm}) shows that the large positive XMCD at the Os $L_{3}$ edge, is the defining aspect of the spectra that come from the anti-parallel nature of the spin and orbital moments in this material. In the case of NpOs$_{2}$, however, the two moments are parallel in agreement with the theory presented in this paper.

In conclusion, we have measured and calculated the x-ray absorption spectra and x-ray magnetic circular dichroism on NpAl$_2$, NpFe$_2$, NpOs$_2$ and PuFe$_2$. The spectra calculated by density functional theory agree well with experimental ones for low values of Hubbard $U$ parameter (approximately 1eV). Using sum rules we have extracted the spin and orbital moments as well as the magnetic dipole contribution of the actinide elements for NpOs$_2$ and PuFe$_2$. We have observed a strong effect of surface treatment on behavior of magnetic domains, preventing some of the systems from reaching magnetic saturation even in these strong magnetic fields. Induced moments on Os atoms in NpOs$_2$ were extracted and, in agreement with theory, the spin and orbital moments are aligned parallel to each other, in contrast to Os embedded in some 3$d$ metals or alloys.

\
\begin{acknowledgments}
We thank P. Voisin and S. Feite for the technical support and the E.S.R.F. safety group, especially P. Colomp. We would like to thank P. Poulopoulos for providing us with the FeCoV:Os sample, to S. Uhle, G. Pagliosa, D. Bouexiere for their technical support at ITU, and to W. G. Stirling for encouragement in the early stages of this project. J. R. and P. M. O. acknowledge support of the Swedish Research Council and the Swedish National Infrastructure for Computing (SNIC). The high purity Np metals required for the fabrication of the compound were made available through a loan agreement between Lawrence Livermore National Laboratory and ITU, in the frame of a collaboration involving LLNL, Los Alamos National Laboratory and the US Department of Energy.
\end{acknowledgments}

\bibliography{NpX2_vfinal}

\end{document}